\documentclass{article}
\usepackage{spconf,amsmath,epsfig}
\usepackage{subfigure}

\title{Modeling and Evaluating Enhancements in Expanding Ring Search Algorithm for Wireless Reactive Protocols}
\name{N. Javaid, A. Bibi, K. Dridi$^{\pounds}$, Z. A. Khan$^{\$}$, S. H. Bouk}

\address{Department of Electrical Engineering, COMSATS, Islamabad, Pakistan. \\
        $^{\pounds}$LISSI, University of Paris-Est Creteil (UPEC), France.\\
        $^{\$}$Faculty of Engineering, Dalhousie University, Halifax, Canada.}
\begin{document}
%
\maketitle

\begin{abstract}
In case of high dynamic topology, reactive routing protocols provide quick convergence by faster route discoveries and route maintenance. Frequent broadcasts reduce routing efficiency in terms of broadcast cost; $B_k$, and expected time cost; $E[t]$. These costs are optimized using different mechanisms. So, we select three reactive routing protocols; Ad-hoc On-demand Distance Vector (AODV), Dynamic Source Routing (DSR), and DYnamic Manet On-demad (DYMO). We model Expanding Ring Search (ERS); an optimization mechanism in the selected protocols to reduce $B_k$ and $E[t]$. A novel contribution of this work is enhancement of default ERS in the protocols to optimize $B_k$ and $E[t]$. Using NS-2, we evaluate and compare default-ERS used by these protocols; AODV-ERS1, DSR-ERS1 and DYMO-ERS1 with enhanced-ERS; AODV-ERS2, DSR-ERS2 and DYMO-ERS2. From modeling and analytical comparison, we deduce that by adjusting Time-To-Live ($TTL$) value of a network, efficient optimizations of $B_k $ and $E[t]$ can be achieved.
\end{abstract}
\vspace{-0.3cm}
\section{Introduction}
\vspace{-0.2cm}
WMhNs provides accessibility to the users in different rates of nodes' mobilities and densities. Routing protocols operated on WMhNs are responsible for calculating efficient end-to-end paths. There are two categories of routing protocols; reactive and proactive, based on their routing behavior. Bandwidth is a critical issue of wireless networks which introduces routing overhead; routing latency and routing load, while establishing end-to-end paths. So, routing protocols are aimed to optimize the \textit{broadcast}; $B_k$ and \textit{expected consumed time}; $E[t]$. Reactive routing protocols such as AODV [1], DSR [2] and DYMO [3] are designed for high mobile networks. These three protocols optimize their routing overhead of blind flooding by using Expanding Ring Search (ERS) algorithm during route discovery; an optimization technique for blind flooding.

In this paper, we model routing overhead of AODV, DSR and DYMO with their original ERS strategies as AODV-ERS1, DSR-ERS1 and DYMO-ERS1. The routing overhead of reactive protocols mainly depends upon $TTL$ and $waiting\_time$ values of ERS. For improving the performance of ERS, we modify the original ERS in AODV, DSR and DYMO as AODV-ERS2, DSR-ERS2 and DYMO-ERS2. Using NS-2, we evaluate performance of selected protocols in different mobilities. The selected performance metrics for the protocols are throughput, E2ED and NRL. These metrics are computed from simulation results to analyze the behavior protocols with their original and enhanced ERS algorithms (ERS1 and ERS2 in AODV, DSR and DYMO).


\vspace{-0.3cm}
\section{Related Work and Motivation}

[4] models trade-offs of ERS algorithm as one of the techniques to minimize inherent overhead of broadcasting in terms of bandwidth, energy, and processing power consumption by analyzing locating time and overhead metrics.

A general analytical framework to study different search sets based on $TTL$ values is presented in [5]. The framework focuses on the problem of finding best search set for ERS schemes in wireless networks, because performance of such schemes largely depends on $TTL$ values of search set. These values are used sequentially to find desired destination. Best possible search set is analytically evaluated in a scenario in which source is at the origin of the circle in network region.



The authors in [6] investigate the effectiveness of this ERS scheme by exactly counting the number of control messages during routing discovery procedure. They select AODV to compare the routing loads in two cases; with and without ERS. The comparison based on simulations presented in [1] provides understanding about effectiveness of ERS in AODV in terms of routing load reduction.


Detailed framework consisting of modeling of routing overhead generated by three widely used proactive routing protocols; DSDV, FSR and OLSR is presented in [7].
%

In the modeled framework of [8], ERS which is a widely used technique to reduce broadcast overhead is used. This technique works by searching successively larger areas in the network centered on the source of broadcast. The authors discuss that network-wide broadcast is initiated only if $L$ (successive searches) fail, so, an optimal $L$ value is needed to minimize the broadcast cost of ERS. They develop a theoretical model to analyze the expected broadcast cost as a function of $L$. Extensive numerical experiments validate their analytical results by considering a large number of random network topologies of different sizes and path lengths. This model concludes that, by tuning the parameter $L$ to the optimum value, broadcast cost can be reduced up to $52\%$ depending on the topology.

Modeling the routing overhead generated by three well known reactive protocols; AODV, DSR and DYMO is presented in our previous work [6]. In this frame work, route discovery and route maintenance routing overhead is modeled in terms of routing packets and latency costs. The framework majorally focus on the optimization perform by ERS algo. to control flooding overhead in selected protocols. We analyze our framework in NS-2.

Comparative to the above mentioned works, we present a frame work for ERS. We select three reactive protocols; AODV, DSR and DYMO like in [6]. These protocols use ERS as a route discovery mechanism, so we model routing overhead of ERS for these protocols naming it ERS1. While, we also enhance ERS1 by tuning number of hops in search diameters between successive waiting times to improve efficiency of the selected protocols. Moreover, we practically evaluate efficiency of ERS1 and ERS2 in AODV, DSR and DYMO in NS-2 with a very high rate of mobility; $30 m/s$ and with different pause times.

\begin{figure}[!h]
\begin{center}
\includegraphics[
height=11 cm,
width=9 cm
]{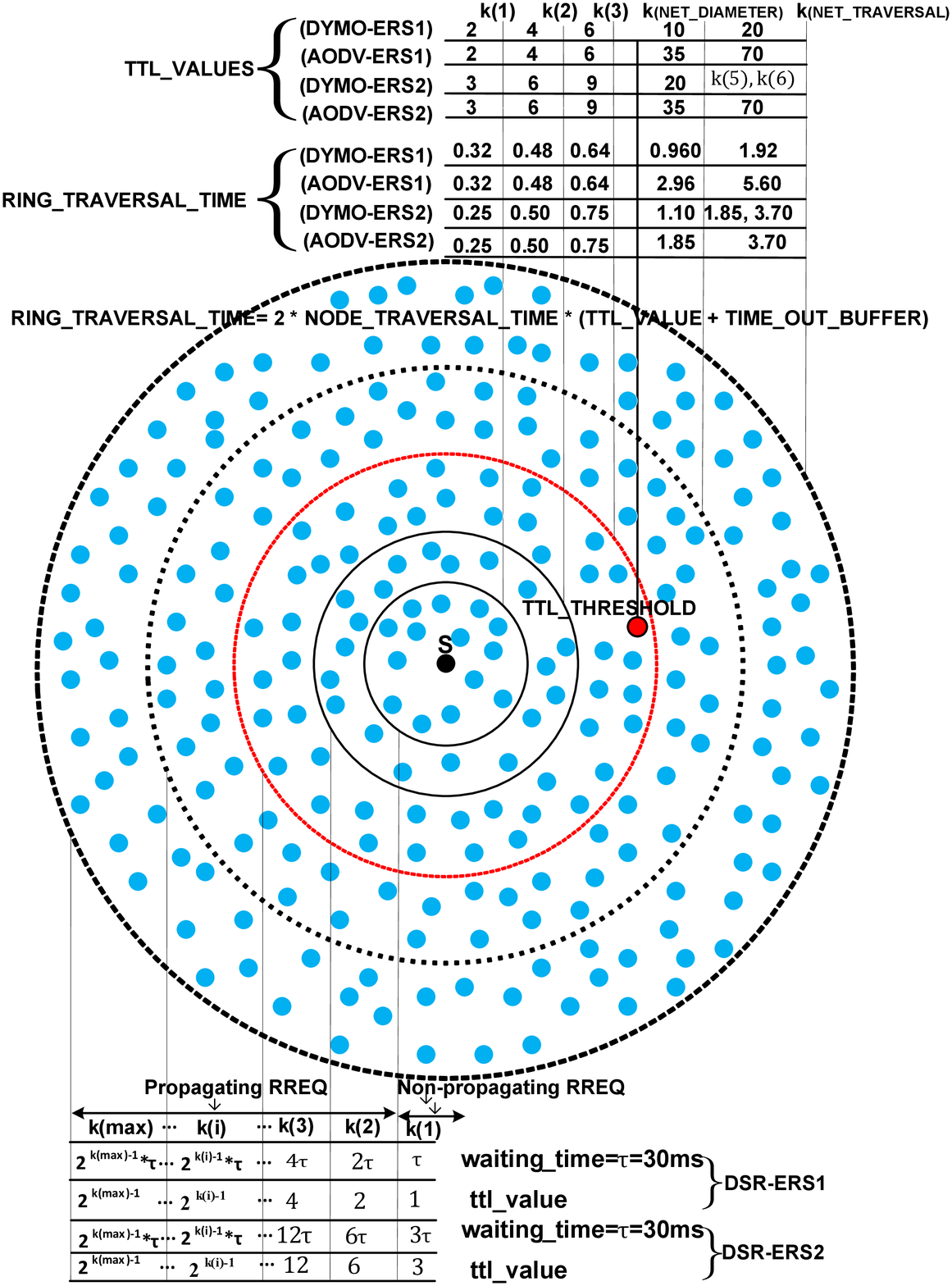}
\end{center}
\vspace{-0.5cm}
\caption{ERS Algorithm used by AODV, DSR, DYMO}
\end{figure}

\vspace{-0.3cm}
\section{Modeling Optimizations by ERS}
\vspace{-0.2cm}
Flooding algorithm is used for exchanging the topological information with every part of a network. In this algorithm, each node can act both as a source and as a router and broadcasts route information to all of its neighbors until destination is reached. A straightforward approach for broadcasting as flooding technique is \textit{blind flooding}; each node is required to rebroadcast the packet whenever it receives the packet for the first time. \textit{Blind flooding} can cause the broadcast storm problem by generating the redundant transmissions.

[8] defines a cost model for network which is to pay for forwarding a query and call it bandwidth cost, $B_k$. This cost specifies the number
of nodes that have to broadcast the query in $N$ (whole network). Let $P_S$ is the stochastic forwarding probability associated with two degrees of connectivity; $d_{avg}$ and $d_f$. Former is the average degree of connectivity with the network and later is the forwarding degree of connectivity with respect to next search set; $k_i$. To accurately measure the flooding cost, we define broadcast cost in whole network; $B_N$ and is given is equation below:

\vspace{-0.3cm}
\small
\begin{eqnarray}
 B_N=
  \begin{cases}
   P_S d_{avg} & if \,\, k_N=1 \\
   P_Sd_{avg}+d_{avg}\displaystyle\sum_{i=1}^{k_N-1}(P_S)^{i+1}\prod_{j=1}^{i}d_f[j]       & otherwise
  \end{cases}
\end{eqnarray}
\normalsize

\begin{table}[h]
\begin {center}
\small
\begin{tabular}{|@{}l@{}|@{}l@{}|@{}l@{}|}
\multicolumn{3}{c}{Table.1. Default and Modified Constants} \\
\hline
\textbf{Constant} & \textbf{Used by}& \textbf{Value(s)} \\ \hline
HELLO\_INTERVAL & AODV, DYMO & 1000ms\\ \hline
LOCAL\_ADD\_TTL &  AODV&2(ERS1),1(ERS2)\\ \hline
MIN\_REPAIR\_TTL&  AODV& Last known hop-count\\ \hline
NET\_DIAMETER& AODV, DYMO &AODV=35, DYMO=\\&&10(ERS1),20(ERS2)\\ \hline
TTL\_START&  AODV, DYMO&2(ERS1),3(ERS2)\\ \hline
TTL\_INCREMENT&   AODV, DYMO&2(ERS1),3(ERS2)\\ \hline
TTL\_THRESHOLD& AODV, DYMO &7(ERS1),9(ERS2)\\ \hline
NODE\_TRAVERSAL& AODV, DYMO &40(ERS1),25(ERS2)\\
\_TIME(ms)&&\\\hline
NET\_TRAVERSAL&AODV, DYMO &AODV=5.6(ERS1),1.1(ERS2)\\
\_TIME(s)&& DYMO=1.92(ERS1),1.1(ERS2)\\\hline
RREQ\_RETRIES&   AODV&2\\ \hline
RREQ\_TRIES& DYMO  &3\\ \hline
TIME\_OUT\_BUFFER&AODV, DYMO &2 \\\hline
DiscoveryHopLimit&  DSR&255 hops\\ \hline
MaxMainRexmt&  DSR&2 retransmision\\ \hline
TAP\_CACHE\_SIZE& DSR  &1024(ERS1),256(ERS2)\\ \hline

\end{tabular}
\end{center}
\end{table}

\vspace{-0.3cm}
\textit{A. Basic ERS:} Many optimizations are proposed to control $B_N$ and ERS [9] is one of them which is implemented in AODV, DSR and DYMO. Flooding is controlled in ERS by limiting Time-To-Live ($TTL$) values in $k_i$. Where $k_i$ denotes the term 'ring' in ERS.


\vspace{-0.3cm}
\small
\begin{eqnarray}
 d_{avg}=
  \begin{cases}
   \displaystyle\sum_{k_i=1}^{k_N}\frac{d_f[k_i]}{k_N} & for \,\, flooding\\
   \displaystyle\sum_{k_i=1}^{M}\frac{d_f[k_i]}{M} & for \,\, ERS
  \end{cases}
\end{eqnarray}
\normalsize

Here, $d_{avg}$ value varies with respect to $k_i$, $k_N$ denotes maximum search diameter of a network and $M$ is the total number of $k_i$ in ERS.

\textit{B. Default ERS:} Basic ERS implemented by existing AODV, DSR, adn DYMO are referred as default ERS for this study.

\textit{B.1. Expected Energy Consumption:} When a limited radius search is started with a radius $k=0$, the respective broadcast cost for this search is $B_k$ which is the number of nodes contained in all the rings up to $k-1$, and $n_i$ denotes the number of nodes in a ring. [8] proposed the following equation:

\vspace{-0.3cm}
\begin{eqnarray}
B_k=1+\displaystyle\sum_{i=1}^{k-1}n_i
\end{eqnarray}

To accurately measure $B_k$, for any ring, $k_i$ in terms of $ttl\_values$, $d_{avg}$, $d_f$, and $P_S$, we present an equation given below:

\vspace{-0.5cm}
\small
\begin{eqnarray}
 \hspace{-1cm}B_{k_i}=
  \begin{cases}
   P_S d_{avg}  & if\,  TTL(k_i)=1 \\ \nonumber\vspace{0.5cm}
   P_S d_{avg}+d_{avg}\displaystyle\sum_{TTL=1}^{TTL(k_i)-1}(P_S)^{TTL+1}\prod_{j=1}^{TTL}d_f[j] & otherwise \,\,\,\,\,(3a)\\
   \end{cases}
\end{eqnarray}
\normalsize

\vspace{-0.5cm}
\begin{eqnarray}
B_M=\sum_{i=1}^{M}B_k(i)
\end{eqnarray}

\textit{B.2. Expected Time Consumption:} [4] modeled expression for expected time in ERS which is obtained by summing all $B_k$ values to maximum ring, $M$ and given in eq.5. Let $T$ denotes timeout after which the next incremented search is initiated, if no reply is received within this $waiting\_time$. $T$ can either be a dynamic value and varies as $k$ varies, or can be a fixed value despite of the search diameter. Hassan \textit{et al.} [8] use fixed $T$ approach, and consider that locating time for each try will vary from $0$ to $T$. They use an average from this range (= T/2). Expected Locating Time is denoted by $E[t]$, as given below:

\vspace{-0.5cm}
\begin{eqnarray}
E[t]=T\displaystyle\sum_{i=1}^{L}(i-1)\times P(i)-LT\displaystyle\sum_{i=1}^{L}P(i)+LT+0.5T
\end{eqnarray}

\vspace{-0.5cm}
\small
\begin{eqnarray}
 E^{DSR}[t]=
  \begin{cases}
  \tau & if \,k_i=1 \\
   \displaystyle\sum_{{k_i}=1}^{M}2^{{k_i}-1}\times \tau  & otherwise
  \end{cases}
\end{eqnarray}
\normalsize

\vspace{-0.5cm}
\small
\begin{eqnarray}
E^{AODV,DYMO}[t]=
  \displaystyle\sum_{{k_i}=1}^{M}\tau_1(TTL(k_i)+\tau_2)
  \end{eqnarray}
\normalsize

Where, $L$ is search threshold. $\tau$ is a constant value for $NonPropagating RREQ$ [2] in DSR with a value of $30ms$ for ERS1 and $90ms$ for ERS2. $E^{AODV,DYMO}[t]$ is shown in Fig. 1 as $RING\_TRAVERSAL\_TIME$. $\tau1$ is $2\times NODE\_TRAVERSAL\_TIME$ and $\tau2$ is $TIME\_OUT\\\_BUFFER$, as are given in Table.1. Eq.7 represents the locating time or $waiting\_time$; $E[t]$ for DSR-ERS1 and DSR-ERS2. $E[t]$'s in $k_i$ of DSR for ERS is portrayed in Fig. 1 and corresponding values are given in Table.2. Eq.8 shows $waiting\_time$ for AODV and DYMO, where, we decrease $NODE\_TRAVERSAL\_TIME$ as $25ms$ (ERS2) from $40ms$ (ERS1) in $\tau1$.


\textit{C. Modified ERS:} In AODV-ERS1, $TTL\_VALUE$ is set to $TTL\_START$ value which is 2 (as mentioned in AODV's RFC[1], then its $TTL\_INCREMENT$ value (=2) is added every time to the previous $TTL\_VALUE$ until a $TTL\_VALUE$ becomes equal to $TTL\_THRESHOLD$, which is equal to 7 (Table.1). For AODV-ERS2, we have changed $TTL\_START$ value to 3, $TTL\_INCREMENT$ value to 3 and $TTL\_THER\\ESHOLD$ value to 9. Moreover, $TTL\_ADD\_LOCAL$ in local link repair is changed to 1 in ERS2. Up to $TTL\_THRESH\\OLD$ route discovery mechanism of DYMO-ERS1 and DYMO-ERS2 are the same as that of AODV-ERS1 and AODV-ERS2, respectively. For $NETWORK\_DIAMETER$, DYMO-ERS1 uses $TTL\_VALUE$ of 10, and for $NETWORK\_TRAVERSAL$ its $TTL\_VALUE$ is 20. In DYMO-ERS2, we have introduce three $TTL\_VALUES$ after $TTL\_THRESHOLD$: (1)$NET\\WORK\_DIAMETER(=20)$, (2)$NETWORK\_TRAVER\\SAL1(=35)$, and (3)$NETWORK\_DIAMETER2 (=75)$. We introduce larger TTL values for dissemination in the network for scalability purpose of DYMO, as its previous ERS (ERS1) is designed for smaller scalabilities.


\begin{figure}[!t]
  \centering
 \subfigure{\includegraphics[height=3 cm,width=6 cm]{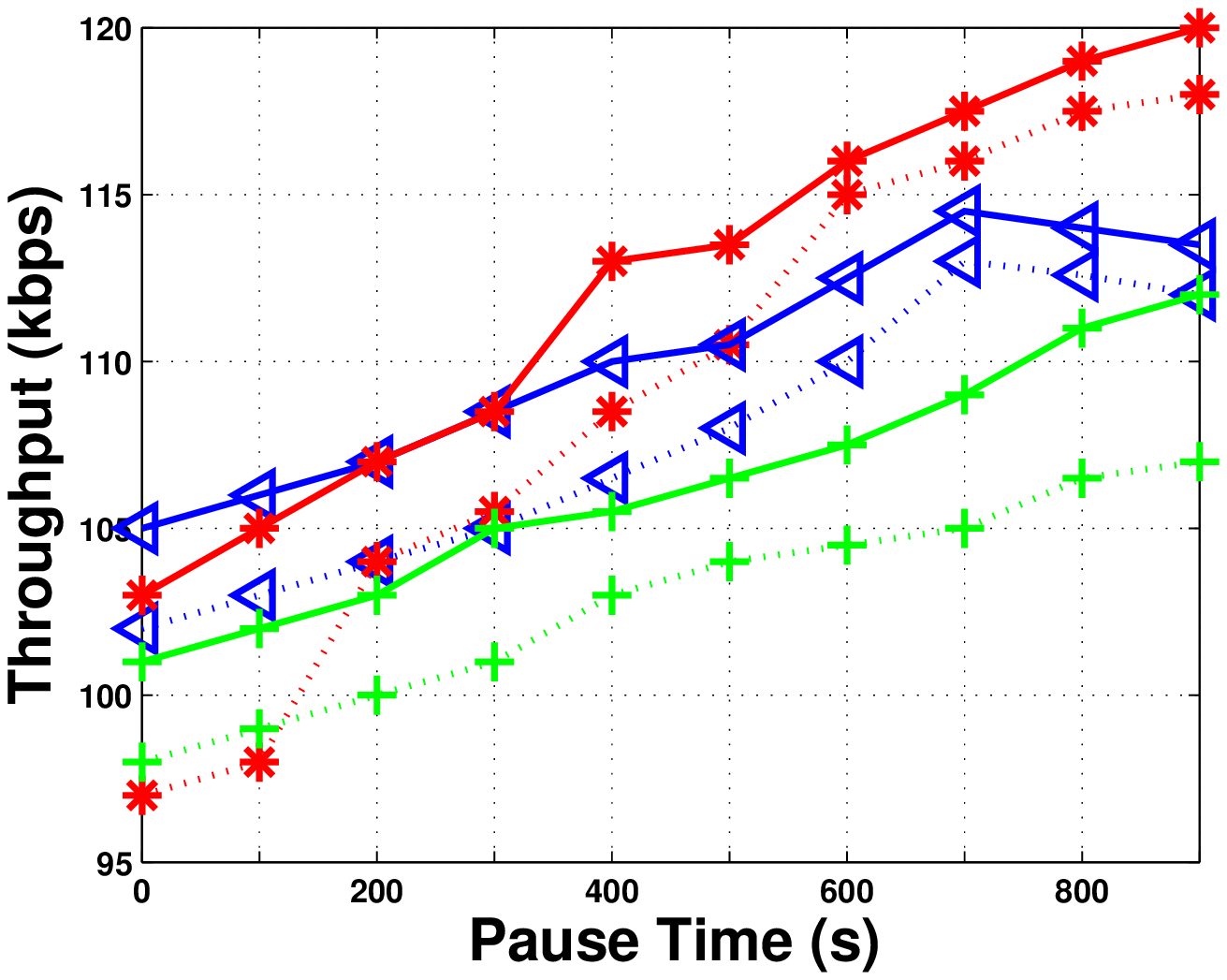}}
 \subfigure{\includegraphics[height=3 cm,width=6 cm]{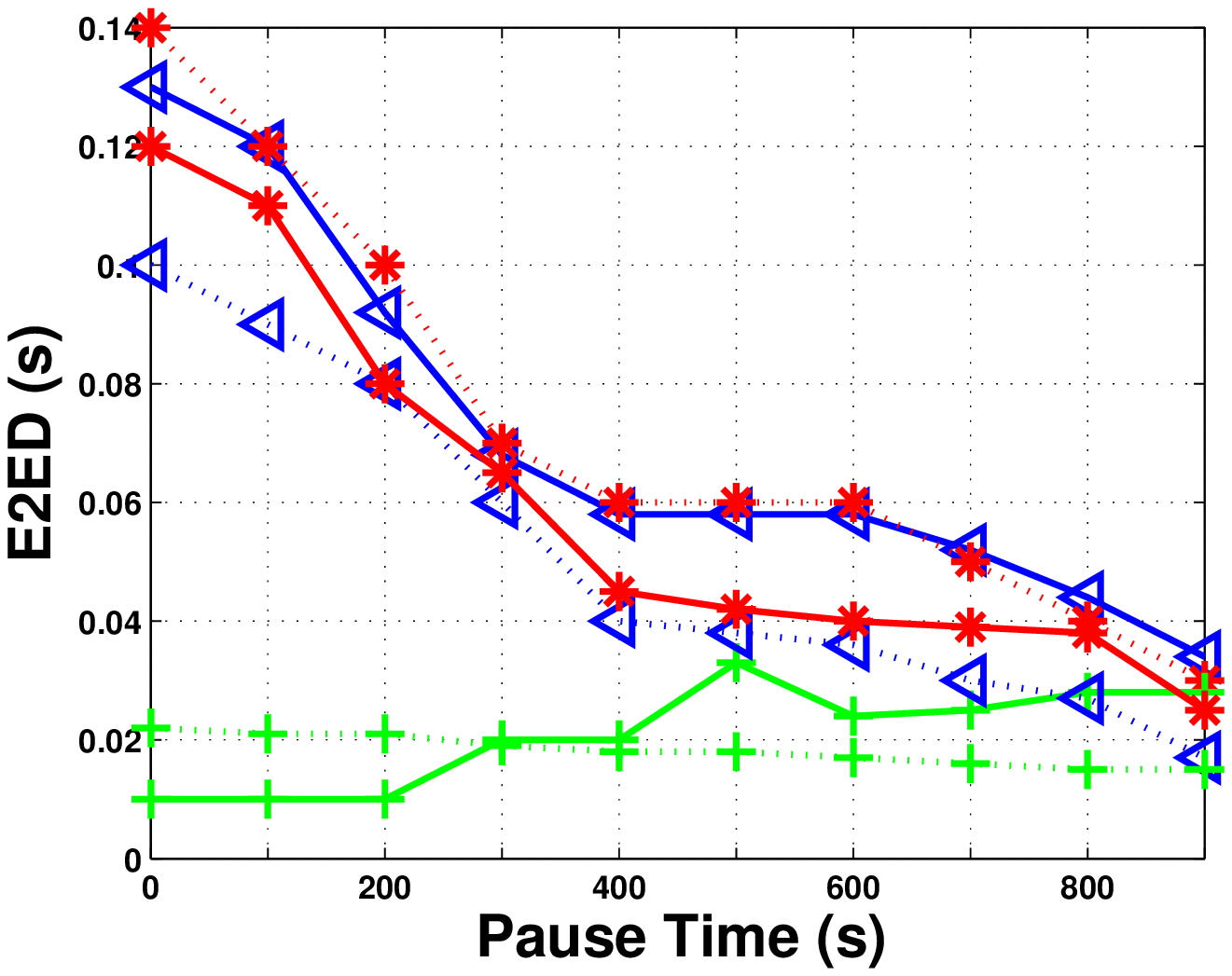}}
  \subfigure{\includegraphics[height=3 cm,width=6 cm]{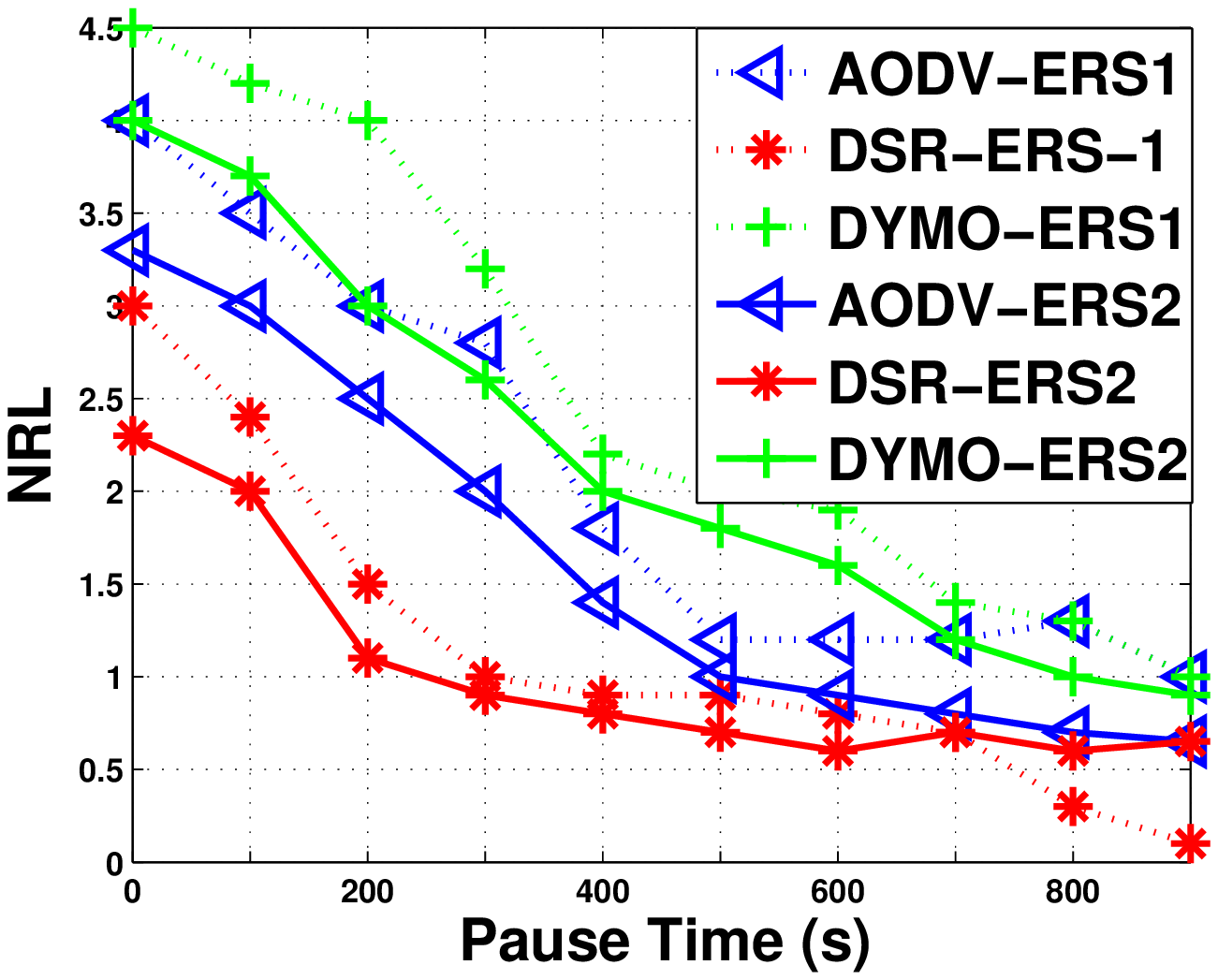}}
 \vspace{-0.3cm}
  \caption{Simulation results to validate the modeling of ERS}
\end{figure}

\vspace{-0.3cm}
\section{Simulations}
To evaluate enhanced ERS with default ERS in the selected protocols, we conduct experiments in NS2. Nodes are dispersed in $1000m\times 1000m$ of network square space having links of a bandwidth of 2Mbps. Continuous Bit Rate (CBR) traffic sources are used with a packet size of $512\,bytes$ and Random Way-Point as a mobility model.


\textit{A. Throughput:} In high mobility, at $0s$, $100s$, $200s$ pause times, AODV-ERS1 attains more throughput than DSR-ERS1. This is because of storage of stale routes in route cache in DSR-ERS1. While AODV-ERS1 and AODV-ERS2 check the route table with valid time and avoid to use the invalid routes from routing table along with a quick repair mechanism for route re-establishment known as local link repair (Fig. 2.a). Whereas, DSR-ERS2 produces high throughput in Fig. 2.a than AODV-ERS1 due to reduction to one fourth from the original size of route cache. This helps to delete older routes, while the initial ring size of DSR-ERS2 of $TTL$ value three make its discovery faster.

The worst performance of DYMO-ERS1 is noticed in Fig. 2.a, among reactive protocols by producing lowest throughput value. The absence of grat. RREPs, local link repair, packet salvaging, route caching, and dissemination of source route information collectively reduce its efficiency. Whereas, due to small value of $RING\_TRAVERSAL\_TIME$ in DYMO-ERS2 increases its throughput as compared to DYMO-ERS1, as, depicted in Fig. 2.a.


\textit{B. E2ED:} In low mobilities, DYMO-ERS1 attains the lowest routing latency because of $TTL(k_1)=1$ value is more suitable for stable networks, as shown in Fig. 2.b. Whereas, DYMO-ERS2 in high mobilities produces lowest delay because $TTL(k_1)=3$ results quick searching.  AODV among reactive protocols attains the highest delay because LLR for link breaks in routes sometimes result in increased path lengths. By increasing $TTL\_START$ and $TTL\_INCREMENT$ values in AODV-ERS2 results more routing latency as compared to AODV-ERS1. At higher mobilities, DSR-ERS1 suffers the higher AE2ED, because stale routes in route caches consumes time of for packet salvaging and route caching. Whereas, incrementing search diameter of $NonPropagating\,RREQ$ and reducing $TAP\_CACHE\_SIZE$ result less routing delay in DSR-ERS2 as compared DSR-ERS1 (Fig. 2.b).

Presence of stale routes in DSR-ERS1 augments routing delay by producing faulty gratuitous RREPs, and thus increases route re-discoveries. In DSR-ERS2, dissemination of larger $TTL$ value of first ring  ($NonPropagating\,RREQ$) exponentially affects the sizes of successive rings, and thus produce less as compared to DSR-ERS1. Moreover, in enhanced DSR, storage of stale routes are avoided due small size of route cache, as a result routing latency is reduced by diminishing route re-discoveries.


\textit{C. NRL:} In the case of different mobilities, due to absence of gratuitous RREPs and route caching, DYMO-ERS1 produces higher routing overhead (Fig.4.e) than all the remaining selected protocol. Whereas, DSR, due to promiscuous listening mode, has the lowest routing load; used for route caching and packet salvaging. By increasing the initial ring size in AODV-ERS2 (from $2$ to $3$, Table.2, $TTL\_VALUES$), DSR-ERS2 and DYMO-ERS2 reduce the routing overhead as compared to ERS1. $ADD\_LOCAL\_TTL$ value is decremented to $1$ from $2$ (Table.1) in AODV-ERS2, this change makes the routing load less during route maintenance (Fig. 2.c).

\vspace{-0.3cm}
\section{Conclusion}
\vspace{-0.2cm}
Routing protocols tackle the issue of broadcasts' routing latency. We first study and model the operations of three protocols; AODV, DSR and DYMO. Modifications in search diameters and waiting\_times can decrease routing overhead. Therefore, we make some enhancement in default ERS in routing protocols naming them AODV-ERS2, DSR-ERS2 and DYMO-ERS2 and compared them with AODV-ERS1, DSR-ERS1 and DYMO-ERS1. A model is constructed for basic and enhanced ERS algorithms. To validate effects of ERS enhancements on routing latencies and broadcast overhead in selected protocols, we conduct simulations in NS-2 using three performance metrics; Throughput, NRL and E2ED. We finally analyze that increasing the search diameter with a suitable waiting\_time not only lessens the routing overhead but also augments the throughput.



\end{document}